\documentstyle[12pt,epsfig]{article}
\author{Luis J. Boya
\footnote{luisjo@posta.unizar.es}\\
Departamento de F\'{\i}sica Te\'orica. Facultad de Ciencias.\\
Universidad de Zaragoza. 50009-Zaragoza, Spain}
\title{Rejection of the Light Quantum: The Dark Side of Niels Bohr
\footnote{† Open Lecture presented at the $4$th Workshop on Rigged
Hilbert Space and Resonances. Jaca,HU, Spain, June $2.001$}}
\begin{document}
\maketitle
\begin{abstract}
	Evidence is recalled of the strong opposition of Niels Bohr, at the time of the
	Old Quantum Theory $1.913 -´25$, to the \emph{Lichtquanten}
	 hypothesis of Einstein. Some
	episodes with H.A. Kramers, J.C. Slater and W. Heisenberg are recollected; Bohr's
	changing point of view is traced back to some philosophical antecedents and to his
	endeavour to deduce quantum results from the Correspondence Principle. Some
	consequences for the future interpretation of Quantum Mechanics, specially to the
	Complementarity Principle, are considered.
\end{abstract}
\vspace{2cm}

\vspace{.5cm}

\newpage

\newcommand{\be}{\begin{equation}}
\newcommand{\ee}{\end{equation}}
\newcommand{\ci}{\cite}
\newcommand{\la}{\label}

\section{BOHR AND THE LIGHT QUANTUM.}

In the Trilogy \ci{1}, the famous $1.913$ paper
		""On the Constitution of Atoms and Molecules"" in three parts by Niels BOHR
		($1.885-1.962$), there are quite a few mentions to the Light Quantum
		Hypothesis \ci{2}.
 For instance \ci{1}, p. 6

"… "...the energy radiation from an atomic system does not take place in the continuous way
assumed in the ordinary electrodynamics but..., on the contrary, takes place in distinctly
separated emissions..."…"

	Later in the same Trilogy Bohr describes the radiation emitted by the atom
	as [ I, p. 172]

		" … "...emission of one of several quanta""

which later turned to be mainly only one. Bohr, however does not use Einstein´s word
\emph{Lichquanten} but Planck´s \emph{energy} quanta.

	So at the very begining Bohr embraced both the failure of classical electrodynamics
	to acount for the atom´s stability, and also the emission by discrete quanta. But
     very soon he changed the perspective: classical electrodynamics was to be mantained,
     indeed used as a guide, quantization limited to matter, and all references to discrete
     emission of (energy) quanta discarded. What brought about this change? We can trace
     back precisely the moment: when Bohr discovers, and tells Rutherford at once in a
     letter \ci{3}, that for large quantum numbers the frequency of the emitted radiation
     coincides with the frequency of the revolution electron, as the clasical theory would
     predict; Bohr had discovered (letter to Rutherford, $21-III-1.913$)

"...the most beautiful analogi [sic] between the old electrodynamics and the considerations
used in my paper"

	This was crucial for Bohr future view of quantum physics, and through Bohr´s
	overwhelming influence to the whole of the physics community:
	the \emph{Correspondence Principle}
	had germinated; indeed, it was criptically expresed already in p. $13-14$
	of the Trilogy. Henceforth Bohr takes it as the guiding principle to discoveries in
	the old quantum theory. Indeed, the original quantum conditions of Planck on
	oscillators, in which Bohr's first atom paper is based, are \emph{rejected}.
	Heilbron and
	Kuhn \ci{4} already realized the contradictions
	between the July, $1.913$ trilogy, and the
	December-$1.913$ Copenhagen communication \ci{5}.
	In the latter, Bohr considers \emph{misleading}
	the original derivation of the quantum rules (kinetic energy
	$T = \nu h /2$, the original
	quantum rule that Bohr borrows from Planck): the
	\emph{correspondence criterion} is elevated
	to the correspondence \emph{principle}.

	In his writings $1.913-1.920$ Bohr steadily gets farther
	and farther away from the photon
	concept, and relying more and more on the classical radiation theory.
	A few quotations
	will suffice \ci{6}:

"...the theory of spectra must be considered in a certain sense as the rational generalization
of the ideas of the usual radiation theory" [$1.918$]

"...On the other hand radiation had to be described by the classical electromagnetic
theory" [$1.921$]

" "[interference phenomena] cannot possibly be understood on the basis of a theory such
as Newton´s. In fact, the picture provided by Einstein … cannot more than give any sort
of explanation of the interference phenomenon" [$1.918$]

	There is  no point in keeping quoting Bohr: all the authors agree on that Bohr
	rejected the corpuscular nature of light. Particularly detailed studies are in
	Pais \ci{7} and Murdoch \ci{8}. Timidly, Pais points out the true reason: the
	correspondence principle clashes head-on with the light quantum concept!

There is some irony in Bohr's writings of the epoch; let us look at two cases. Sommerfeld
in M\"unchen was developing a quasi-consistent quantization system of rules for atomic
systems, and was not happy with the correspondence principle, as "
\emph{"...etwas Fremdartigen"}
to the theory, a ""magic wand, which operates only in Kobenhavm"". Bohr, of course,
disliked Sommerfeld approach:

"...in every fine point that came up Sommerfeld was wrong" \ci{9}.

	The irony was, of course, that the difficult Sommerfeld
	\emph{Feinstruckturformel} was one
	of big successes of the Old Quantum Theory, an issue in which him, Bohr, failed
	(he did  apply relativity to the
	\emph{circular} orbits only). Sommerfeld never got the
	Nobel prize, to the surprise of many (including A. Pais, an admirer of Bohr).
	Actually, Bohr never recommended Sommerfeld  to the Swedish Academy, whereas most
	of his recommendees did get the Prize (for a full list see \ci{10}; \ci{11}).

	In a comment on Einstein´s \emph{Lichquantum}, Bohr ($1.920$) had this to say

"...the Light Quantum hypothesis is so formal, that even Einstein himself shrouded
       it of mysticism, talking about the \emph{Gespensterfeld} to guide the "photons"" "

	The irony here is double: first, Bohr himself, in the ill-fated BKS paper (which
	we shall comment later) introduced the ""guiding field" in terms of virtual
	oscillators, altough he borrows the idea from Slater;
	and secondly, Einstein´s $1.920$
	(unpublished, see \ci{7}, p. $287$) idea was later, in the hands of Schr\"odinger,
	instrumental in founding the wave mechanics, which Bohr embraced! (see also later)
	because after all it came to the rescue of his beloved classical e.m. theory.
	The \emph{Gespensterfeld}, in the hands of Max Born,
	became the basis for the probability
	interpretation, today universally accepted, also by Bohr, as part of \emph{his}
	complementarity point of view. Of course, complementarity and the Born interpretation
	are at odds which each other. It had to be a philosopher (Karl Popper, \ci{12}), coming
	from outside, who pointed out the contradiction.

\section{KRAMERS vs. KRAMERS.}
It will be prophilactic to
describe now the episodes in which  Bohr clashed with people on the "photon" issue.
The \emph{first victim} was (Hans) A. Kramers ($1.894-1.952$),
Bohr's first collaborator, and a "
"yes man" for Bohr, as Slater later put it. Now the historian-physicists A.Pais and
M. Dresden talk ( I resume from \ci{13} and \ci{14}):

In Spring, $1.921$ "The indications are that Kramers had obtained the conservation law
description of the Compton effect, using the photon description explicitely...…Bohr
would object violently to the publication of these results. He and Kramers engaged
inmediately in a series of daily "no holds barred"
\footnote{The spanish sentence \emph{a calz\'on quitado} is more vivid.}
arguments about the photon...In
these discussions...Kramers...was simply grounded down by Bohr...…
After these discussions, which left Kramers exhausted, depressed, and let down, [he] got
sick and spent some time in the hospital. During his stay in the hospital, Kramers gave
up the photon notion...altogether. Instead, he soon became violently opposed to the photon
notions, and never let an opportunity pass by to criticize or even ridicule the concept.
He disposed of most of his early calculations, but inadvertedly left a few, early rough
notes  [which are extant, fortunately, in  Kramers's family's hands. LJB]

	The Bohr-Kramers discussions are in spring, $1.921$. That year Bohr was overworked
	with the building a new institute, and had to renounce to participate in the
	Solvay meeting. But a glance of the spiritual turmoil is felt, I think, in the
	following revealing letter to Ehrenfest \ci{15}

"You have no idea how much your friendship means to me. Especially at a time when I
almost feel as a criminal in relation to all kinds of people here and elsewhere""

	As the incident with Kramers is an important one, let us recall that already in
	his $1.967$ book, ter Haar said \ci{16}

"Debye ($1.923$) developed the theory of the Compton effect as did Kramers, who was
persuaded by Bohr not to publish..."…"

	Further evidence comes from Jost, Pauli, O. Klein,  Hugenholtz, Casimir,
	and others. It is interesting to read Kramers
	\emph{abjuration}, which reminds me
	very much of the self-indicted declarations of the comunists, in Russia or China:

"The theory of Light Quanta might be compared with a medicine which will cause the
disease to vanish, but kills the patient. The fact must be emphasized that this theory
in no way has sprung from Bohr theory, to say nothing of its being a necessary
consequence of it"\footnote{ Kramers-Holst book, ($1.923$) \ci{17}}"

	Were it not because that title belongs to Leon Rosenfeld, one would call Kramers
	"
	\emph{"Plus royalist que le Roi"}
	\footnote{"Again, the spanish expression
	\emph{M\'as papista que el Papa} is better.}
	with respect to Bohr. One should also feel sorry
	for Kramers; the fact that he was a man with poor health alleviates the burden.
	His argument to reject the photon was, it seems, that it was not invented by
	God (\emph{alias} Niels Bohr).
	The irony is that precisely it is in the Bohr atom that the
	photon concepts enters in the most natural way. Many modern books describing the
	Bohr atom mention expressly the emission of a photon, when a valence electron
	makes a transition, as if it was part of the original hypothesis of Bohr!

	Kramers left Bohr in $1.926$ to a distinguished career in his native Holland.

\section{J. C. SLATER.}

The \emph{next victim} is the brilliant american physicist
		John Clark Slater ($1.900-1976$).
		I show here some evidence gathered from \ci{18},
		Pais again \ci{7}, and Slater Recollections \ci{19}. Slater conceived around
		November $1.923$ in England a theory to reconcile the apparent wavelike
		properties of light (interference and diffraction) with the corpuscular
		photon of Einstein. Slater had the excited atom to "communicate""with
		other atoms by virtue of some "virtual field""before emitting a photon:

"As soon as the atom reaches a stationary state the [virtual] field is set up...…
containing all the frecuencies the atom can radiate. These fields determine the
probabily of emission of a quantum...Finally a quantum of someone of the frequencies will
be emitted, [and] the radiation will cease. Meanwhile the quantum is travelling...until
it is absorbed...…
 	
 	As soon as I discussed [this] with Bohr and Kramers [Dec-$1.923$]
 	I found...to my
 	consternation that they completely refused to admit the real existence of the
 	photons
 	\footnote{Slater wrote this much later, when already the
 	\emph{Lichtquantum} of Einstein was universally
	called the photon, name due to G.N. Lewis, ($1.926$).}.
 	It had never ocurred  to me that they would object to what seemed
 	like so obvious a deduction from many types of experiments...They grudgingly
 	allowed me to send a Note to \emph{Nature}
 	indicating that my original idea had included
 	the real existence of the photons, but that I had given that up at their
 	instigation""[Slater´s abjuration, see soon below. LJB]

In a letter to van der W\"arden \ci{18} Slater has this to say ($1.964$)

"...Bohr and Kramers opposed this view [of the photon] so vigorously that I saw the only
way to keep peace and get the main part of the suggestion [the virtual oscillators]
published was to go along with them..."

	Before relating Slater´s abjuration, let us comment briefly the ill-fated
	$1.924$ Bohr-Kramers-Slater paper. It was written entirely by Bohr and Kramers,
	while Slater was kept locked up in another room. The paper has $18$ pages with a
	single formula,
	$E = h\nu$. Bohr gave up conservation of energy lest the photon survive.
	Einstein, Heisenberg and Pauli opposed strongly; of course, energy conservation
	was vindicated in both sides of the Atlantic pretty soon \ci{20}. Did then Bohr
	accepted the photon ? No! But let us first listen to Slater confession:

"...it seems possible to build up a more adequate  picture of optical phenomena than
has previously existed, by associating the
\emph{essentially continuous} radiation field with
the continuous existence of stationary states, and the discontinuous changes of energy
and momentum with the discontinuous transitions from one state to another"
[my italics. LJB] \footnote{Letter sent to Nature, $28-I-1.924$}

	In other words, Slater renounces to the photon...Some scattered comments convey
	more the spirit of the abdication:

"I have finally become convinced that the way they [Bohr and Kramers] want things,
without the little lump carried along the waves...is better...I am going to have a chance
at least to suggest changes""\ci{21}

Eventually Slater received an apology from Bohr: "I had bad conscience about you when
in Copenhagen..." J. C. Slater went on to another distinguished career in physics in the East
Cost, being instrumental in buiding the theoretical physics in the U.S., as the
\emph{transition}
person from the meteoric start of W. Gibbs to the preeminence with J. R. Oppenheimer, etc.

	After the Compton-Simon and Bothe-Geiger experiments  \ci{20} (spring $1.925$),
	Bohr renounced to statistical-only conservation of energy and momentum, but he
	does \emph{not} embrace the photon.
	Heisenberg says he believed around early $1.925$ Bohr
	was the only notorious physicist unbeliever about photons. Bohr's conclusion instead
	was: one should renounce to a spacetime description of physical atomic phenomena;
	he took cold confort in another enigma at the time, the discovery of the
	(Townsend-) Ramsauer effect (anomalous low energy scattering of electrons by
	noble gases), to delve in (meta-)physical thoughts about non space-time
	descriptibility of microscopic phenomena; the irrationality so apparent in
	many of Bohr's statements in $1.927-30$, somewhat attenuated afterwards, take their
	roots here.

\section{SCHR\"ODINGER and HEISENBERG.}
In September, $1.926$ E. Schr\"odinger ($1.887-1.96$1)
	joined W. Heisenberg ($1.901-1.976$) and N. Bohr in Copenhagen. We shall learn more
	of the way Bohr dealt with opponents. The report  is by Heisenberg \ci{22}:

"The discussions between Bohr and Schr\"odinger began already at the train station and
were continued each day from early morning til late at night...And Bohr...now appeared to
me almost as an unrelentic fanatic, who was not prepared to make a single concession to
his discussion partner...So the discussion continued for many hours throughout day and night
without a consensus being reached. After a couple of days, Schr\"odinger fell ill…
...He had to stay in bed with a feverish cold. Mrs. Bohr nursed him and brought tea and cakes,
but Niels Bohr sat on the bedside and spoke earnestly to Schr\"odinger..."

	At issue was the wave-like interpretation of the wave mechanics, no so much
	the photon; but I bring this incident to sharpen the reader´s ideas on Bohr.
	Schr\"odinger later reported to a friend how was he [S.] astonished by Bohr´s
	happy coexistence with contradictions bordering the irrational; also

"...There will hardly again be a man who...in his sphere of work is honored almost like
a demigod by the whole world and who yet remains...rather shy and diffident like a theology
student...

[Bohr] talks  often for minutes almost in a dreamlike, visionary and really quite unclear
manner..."\ci{23}

	We are approaching the climax now. When Schr\"odinger left, the two men, Bohr
	and Heisenberg, embarked, through a socratic dialogue, to set up the conceptual
	foundations of quantum mechanics, what soon will become the
	\underline{Copenhagen
	interpretation}; but the starting points of them were very different: Bohr took
	the correspondence principle as a guide, and had already extracted the (unwarranted ?)
	conclusion that

		" "Every description of natural processes must be
	based on ideas which have been introduced and defined
 by the classical theory" ($1.923$; \ci{7}, p. $300$)

	Bohr would take this phrase almost literally to
	\emph{hold also in the new interpretation}
	(see later, Como conference report). As for Heisenberg,

		" "I dislike this  [Bohr´s] approach. I want to start from
the fact that quantum mechanics [H. here means the G\"ottingen matrix theory as started by
him, elaborated by Dirac and in the
\emph{Dreim\"annerarbeit}, and sharpened in the transformation
theory of Jordan and Dirac; he is more explicit in \ci{24}]...already imposed a unique physical
interpretation...so...we no longer had any freedom with...interpretation""\ci{7}, p. $303$).

	In correspondence of H. with Pauli, which I shall not reproduce here, it is
	clear the main fighting point: Bohr wanted to include waves, and to make a
	blend of  Schr\"odinger wave mechanics, but of course not his [S.] interpretation,
	whereas H. insisted in the particle point of view, with Max Born´s probability
	interpretation for the matrix elements, in particular, for the wave function itself;
	(von Neumann ($1.929$) would definitely clarify this, showing that Heisenberg was
	using a \emph{energy} representation, Schr\"odinger a
	\emph{coordinate} representation, equally
	valid, and equivalent mathematically). That much is perfectly clear in Heisenberg´s
	original writing of the
	\emph{Umbestimmheit} paper (March, $1.927$). When Bohr came back of
	his skiing holiday in Norway, he corrected rightly a small mistake in the paper of
	Heisenberg (he had made a similar error in his oral examination for the Ph. D.,
	which nearly costed him the degree: Wien was reluctant, but Sommerfeld came to
	his rescue); then the fight between the two men was very acute, as witnessed by
	Oscar Klein; as described by Heisenberg,
	
"Bohr tried to explain [my paper] was not right and I shouldn't publish the paper.
I...ended by my breaking out in tears because I just couldn't stand this pressure from
Bohr´...So it was very disagreable [for] a short period of perhaps ten days or so in
which we really disagreed rather strongly...[but after] we agreed that the paper could be
published if it was improved on these points..." …"
[ A. Pais, \ci{7}, p. $308$]

	The published version of the uncertainty paper contains, as a \emph{Nachtrag},
	Heisenberg´s abdication:

"After submission of this work,...Bohr has pointed out that I have overlooked
essential features...The uncertainty in an observation depends not only on the
ocurrrence of discontinuities, but also directly on the requirement that...measurements
are to be made...as in particle theory or...as in
\emph{wave theory}...For permission to mention his
new research...I was privileged to learn...at his genesis, I owe my heartfelt thanks to
Professor Bohr" \ci{25}" (my italics. LJB)

	The abdication of Heisenberg was a full
	\emph{volte face}, same as Kramers had experienced
	six years before. The phrase "Kopenhagener Geist der Quantentheorie" is
	Heisenberg's own \ci{26}.

To fully analyze how Bohr passed from \emph{Correspondence}
to \emph{Complementarity} is beyond our
purpose.  In a nutshell, Bohr kept from the old principle about the survival of classical
radiation theory (instead of, as I think it should be, deduce the classical wave behaviour
as statistical averaging of the individual quantum particles; but I shall not argue on this),
the coexistence of particles and waves. Of course, an element of irrationality creeps in,
but Bohr was happy with it \ci{27}, as already Schr\"odinger noticed; see this other testimony,
by Bohr:

"...our endeavor is, by means of a suitably limited use of
mechanical and electromagnetic concepts, to obtain a statistical description of the
atomic phenomena that appears as a rational generalization of the classical physical
theories, in spite of the fact that the quantum of action from their point of view must
be considered as an irrationality" ($1.933$" \ci{28}.

	Interpretation: Bohr invites us to study the quanta as a
	\emph{rational} generalization
	of the classical theory, inspite of the fact that the quantum, from the very
	clasical point of view, is seen as an
	\emph{irrationality}. If this is intelligible,
	\emph{que venga Dios y lo vea}.

 I find this hard to swallow, to say the least, and I think Einstein, Schr\"odinger, Planck,
 and others, who opposed to the Copenhagen interpretation, did that because they were
 unable to digest this. Einstein is very explicit: "I never understood what Bohr means
 by the complementarity principle, inspite of a careful study of it". If the first
 intelligence of the XX Century is unable to grasp the meaning of complementarity, the
 odds are...is un-understable; and perhaps this is what Bohr really had in mind. "We must
 understand, that there is nothing to be understood". Einstein did not \emph{accept} the
 probability interpretation, but did not
 \emph{understand} complementarity; there is a world of
 difference...…

	There some other episodes of rudeness of Bohr, with Landau and with Brioullin,
	for example, which would end to delineate the dark side of Bohr I'm showing; I
	just refer to the cartoon by G. Gamow with respect to Bohr´s reaction to a paper
	by Landau and Peierls ($1.931$) \ci{29}

\section{CONCLUSIONS.} 		

\begin{enumerate}

\item Bohr started using Planck's and Einstein's hypothesis to explain the atom and its
radiation.
\item Very soon, before the first part of the Trilogy is published, he discovers
	that for large quantum number $n$ the radiation and the rotation frequencies
	coincide; it acts like a \emph{revelation} for him.
\item He next substitutes the quantum conditions by the "correspondence principle",
	if dressed with the \emph{adiabatic principle} of Ehrenfest (this is very clear in the
	"Tetralogy" \ci{30})
\item As a consequence, the \emph{classical radiation}
	theory is set up as a \emph{referent} for
	the new discoveries; the little original love for the energy quantum
	dissapears completely. Later Bohr was the staunchest enemy of Einstein´s photon
	concept, to the point of ridicule.
\item Part of the classical heritage is the necessity of speaking of all the
	concepts, \emph{even the quantum ones}, with a classical lenguage; at some point Bohr
	even said. "There is NO quantum world".
\item There is strong competition between the M\"unchen school of Sommerfeld and
	Bohr's in Copenhagen; \emph{each spurns the other}.
\item After the Compton effect, he still is antiphoton, but starts to develop an
	irrational attitude towards the quantum, because coexistence of classical pictures,
	which he wants to mantain at any cost, and the hard reality of genuine quantum
	processes.
\item The advent of quantum mechanics (in matrix form; Heisenberg, July $1.925$) caught
	Bohr \emph{off} guard; he always tought
	of making himself a contribution to its discovery.
\item He became a great fan of Heisenberg, to whom he considered as a kind of Messiah.
\item There is no evidence Bohr ever went through the intricacies of matrix mechanics;
	but he salutates the wave mechanics of Schr\"odinger (January, $1.926$), and after the
	fall-$26$ discussions with Schr. and Hei., embarks himself in a crusade for THE right
	interpretation of Quantum Mechanics. He thinks he achieves this in the snow in
	Norway (February, $1.927$).
\item He brainswashed Heisenberg, as had done with Kramers and Slater, forcing him to
	modify his paper; H. became thoroughly convinced.
\item His climax occurs at the time of the Como conference (Volta Death Centenary,
	$16$ September $1.927$). I cannot refrain repeating some phrases, which reflect
	Bohr's fidelity to his concepts: \ci{31}
"...nuestra interpretaci\'on de los datos experimentales se apoya de manera fundamental
en los conceptos cl\'asicos""(p. $98$).

"Este postulado [the postulate Bohr calls quantum \ci{32}] atribuye una...individualidad
a todo proceso at\'omico e implica la renuncia a una coordinaci\'on causal de los procesos
at\'omicos en el espacio y en tiempo""(p.$99$).

"Por lo que se refiere a la luz, su propagaci\'on en el espacio y en el tiempo queda descrita
de manera satisfactoria por la teor\'{\i}a electromagn\'etica [read: Bohr does still not accept
¡in the  fall 1.927! the propagation of photons]...para llegar a una expresi\'on exacta de
la conservaci\'on de la energ\'{\i}a...es preciso recurrir a la idea del fot\'on de Einstein
[ notice how reluctantly introduces Bohr the photon]...Esta situaci\'on muestra con claridad
la imposibilidad de mantener una descripci\'on causal y espacio-temporal de los fen\'omenos
luminosos...." ( p. $101$).

\item The overwhelming personality of Bohr's wins everybody in the young generation
	(Pauli, Dirac, Landau, even Max Born) to his point of view. Heisenberg will be the
	new prophet:

			" "Since the conclusive studies of Bohr in $1.927$
there have been no essential changes in these [fundamental quantum] principles"" [26, Pref.]

\item Nevertheless, practitioners of Q. M. just \emph{ignore} complementarity. Wigner
was not impressed at all in Como; the books by Dirac and Landau do not mention it.
\end{enumerate}
	
	The degree of brainwashing by Bohr in all of  us is remarkable (I learned the
	expression from Murray Gell-Mann). For example, in the "Welches Weg?""experiments,
	either the two slit screens or Mach-Zehnder interferometer, people say (excuse me
	for not quoting!) that there is \emph{complementarity} between waves or particles bzw.
	according the photon runs through both paths (that is, we mantain coherence) or only
	by one (incoherence, meaning we know which path). But this is as saying that
	interferences are ondulatory, but diffraction is corpuscular!! The photon, a particle,
	is quantic, so his path does not exist if it is not observed. The Feynman´s path
	integral formalism is just perfect to explain the "propension""of the particle
	(photon or Mach truck !) to explore \emph{all} paths.

\section*{Acknowledgements} \nonumber
I thanks listeners in several places for patience and questions.


\begin{thebibliography}{40}
\bibitem{10} See \ci{7}, p. 210. Most of Bohr's proposed people eventually got
the Prize. The case of Sommefeld is flagrant. A. Pais again: "I belong to those who
regret...that Sommerfeld´s work was never sufficiently recognized by the Nobel Committee.
He was nominated every year but one from $1.917$ until (at least) $1.937$". \ci{7}, p. 214
\bibitem{11} Bohr also opposed to the modern Quantum Electrodynamics
(I do not know why),  with the result that Tomonaga, Schwinger and Feynman received the
Nobel Price ($1.965$) after Bohr died. Y. NEEMAN and Y. KIRSH: The Particle Hunters.
Cambridge  U.P., 1.986, p. 78 of the Spanish Ed., Gedisa 1.988: "
"En 1.965 Feynman, Schwinger y Tomonaga compartieron el Premio Nobel de F\'{\i}sica.
Podr\'{\i}an haberlo recibido varios a\~nos antes, pero tambi\'en Niels Bohr ve\'{\i}a con recelo
la nueva teor\'{\i}a y su actitud negativa disuadi\'o al Comit\'e Nobel de reconocer la obra de
aquellos cient\'{\i}ficos.
	La decisi\'on de otorgar el Premio Nobel a los autores de la Q.E.D. fue adoptada
	s\'olo despu\'es de la muerte de Bohr ($1.962$)""
\bibitem{12} K. POPPER: ""Quantum Theory and the Schism in Physics":
Post Scriptum III to "Logik der Forschung", Spanish Ed., Tecnos 1.992, p. 33: "
La conexi\'on que hizo Bohr del <<principio de complementariedad>> con el <<dualismo
de part\'{\i}culas y ondas>> se vino abajo cuando se acept\'o la interpretaci\'on de Born
del cuadrado de la amplitud de onda como una probabilidad para hallar la part\'{\i}cula.
Porque esto significaba, en realidad, la aceptaci\'on que la interpretaci\'on corpuscular
era la b\'asica".

It was this very clear point of Popper which launched me into this historical revision [LJB].
\bibitem{31} The Como lecture is reproduced in "Atomtheorie und
 Naturbeschreibung"; I'm using the spanish. trasl., Alianza
 Universidad 1.988
\end{thebibliography}
\end{document}